\documentclass[copyright,creativecommons]{eptcs}
\providecommand{\event}{GandALF 2014}  % Name of the event you are submitting to
\usepackage[T1]{fontenc}
\usepackage[utf8]{inputenc}
\usepackage[english]{babel}
\usepackage{amsmath,amssymb}
\usepackage{bbm}
\usepackage[thmmarks]{ntheorem}
\usepackage{enumerate}
\usepackage{tikz}
\usetikzlibrary{shapes,automata,trees,backgrounds,arrows}

\input{style/styling-and-macros.sty}
\proofstrue
\appendixfalse

\title{Synthesis of Deterministic Top-down Tree Transducers from Automatic Tree Relations}
\author{
Christof L\"oding \qquad\qquad Sarah Winter
\institute{Lehrstuhl f\"ur Informatik 7, RWTH Aachen, Germany}
\email{$\{$loeding,winter$\}$@automata.rwth-aachen.de}
}
\def\titlerunning{Synthesis of Deterministic Top-down Tree Transducers from Automatic Tree Relations}
\def\authorrunning{Christof L\"oding and Sarah Winter}

\begin{document}
\maketitle

\begin{abstract}
We consider the synthesis of deterministic tree transducers from automaton definable specifications, given as binary relations, over finite trees.
We consider the case of specifications that are deterministic top-down tree automatic, meaning the specification is recognizable by a deterministic top-down tree automaton that reads the two given trees synchronously in parallel.
In this setting we study tree transducers that are allowed to have either bounded delay or arbitrary delay.
Delay is caused whenever the transducer reads a symbol from the input tree but does not produce output.
We provide decision procedures for both bounded and arbitrary delay that yield deterministic top-down tree transducers which realize the specification for valid input trees.
Similar to the case of relations over words, we use two-player games to obtain our results.
\end{abstract}

\section{Introduction}
\label{sec:1}
The synthesis problem asks, given a specification that relates
possible inputs to allowed outputs, whether there is a program
realizing the specification, and if so, construct one. This problem
setting originates from Church's synthesis problem \cite{church1962logic} which was
already posed in 1957. Church considers the case where the input is an
infinite bit sequence that has to be transformed, bit by bit, into an
infinite bit sequence.  The synthesis problem is then to decide
whether there is a circuit which realizes the given input/output
specification, and construct one if possible.  A related notion is the
one of uniformization of a (binary) relation, which is a function that
selects for each element of the domain of the relation an element in
its image. The synthesis problem asks for effective uniformization by
functions that can be implemented in a specific way.

Specifications are usually written in some logical formalism, while
the uniformization, in particular in the synthesis setting, is
required to be implemented by some kind of device. Since many logics
can be translated into automata, which can also serve as
implementations of a uniformization, it is natural to study
uniformization problems in automata theory. Relations (or
specifications) can be defined using automata with two input tapes,
and uniformizations can be realized by transducers, that is, automata
with output. 

A first uniformization result in such a setting has been obtained by
Büchi and Landweber in \cite{buechi}, who showed that for specifications over
infinite words in monadic second-order logic, it is decidable whether
they have a uniformization by a synchronous transducer (that outputs
one symbol for each input letter). The specifications considered in
\cite{buechi} can be translated into finite automata that read the two input
words synchronously. Such relations are referred to as automatic
relations over finite words, and as $\omega$-automatic relations over
infinite words. 

The result of Büchi and Landweber has been extended to transducers
with delay, that is, transducers that have the possibility to produce
empty output in some transitions.  For a bounded delay decidability
was shown in \cite{hosch1972finite}, and for an unbounded delay in \cite{holtmann2010degrees}. In the case of
finite words, it was shown in \cite{uniformizationAutomataTheory} that it is decidable whether an
automatic relation has a uniformization by a deterministic
subsequential transducer, that is, a transducer that can output finite
words on each transition.

Our aim is to study these uniformization questions for relations over
trees. Tree automata are used in many fields, for example as tool for
analyzing and manipulating rewrite systems or XML Schema languages
(see \cite{tata2007}). Tree transformations that are realized by finite
tree transducers thus become interesting in the setting of
translations from one document scheme into another \cite{MiloSV03}.
There are already some uniformization results for tree relations. For
example, in \cite{Engelfriet78} it is shown that each relation that
can be defined by a nondeterministic top-down tree transducer, has a
uniformization by a deterministic top-down tree transducer with
regular lookahead. However, these results focus on the existence of a
uniformization for each relation that can be specified by the
considered model. In contrast to that, we are interested in the
corresponding decision problem. More precisely, for a class
$\mathcal{C}$ of tree relations and a class $\mathcal{F}$ of functions
over trees, we are interested in a procedure that decides whether a
given relation from $\mathcal{C}$ has a uniformization in
$\mathcal{F}$.

In this paper we start the investigation of such questions in the rich
landscape of tree automaton and tree transducer models. We study
uniformization of automatic tree relations over finite trees by
deterministic top-down tree transducers. We distinguish between two
variants of uniformization. In the first setting, we do not require
that a transducer validates whether an input tree is part of the
domain of the given specification. We allow a transducer to behave
arbitrarily on invalid input trees. In the second setting, the desired
transducer has to reject invalid input trees. We speak of
uniformization without resp.\ with input validation. For uniformization
without input validation, we consider the case that the transducer
defining a uniformization has no restrictions. In particular the
transducer is allowed to skip an unbounded number of output symbols
thereby introducing delay. We show that it is decidable whether a
given relation has a uniformization by a top-down tree transducer, and
if possible construct one. For uniformization with input validation,
we see that this variant is more complex than uniformization without
input validation.  We show decidability in case that a transducer
realizing a uniformization, synchronously produces one output symbol
per read input symbol.  

The paper is structured as follows. First, we fix some basic
definitions and terminology. Then, in Section \ref{sec:3} and Section \ref{sec:4}, we
consider uniformization by top-down tree transducers without input
validation that have bounded delay and unbounded delay,
respectively. In Section \ref{sec:5}, we briefly consider the case of
uniformization with input validation.

\section{Preliminaries}
\label{sec:2}
The set of natural numbers containing zero is denoted by $\mathbbm{N}$.
For a set $S$, the powerset of $S$ is denoted by $2^S$.
An \textit{alphabet} $\Sigma$ is a finite non-empty set of letters.
A finite \textit{word} is a finite sequence of letters.
The set of all finite words over $\Sigma$ is denoted by $\Sigma^*$.
The length of a word $w \in \Sigma^*$ is denoted by $|w|$, the empty word is denoted by $\varepsilon$.
For $w = a_1\dots a_n \in \Sigma^*$ for some $n \in \mathbbm{N}$ and $a_1,\dots,a_n \in \Sigma$, let $w[i]$ denote the $i$th letter of $w$, i.e., $w[i] = a_i$.
Furthermore, let $w[i,j]$ denote the infix from the $i$th to the $j$th letter of $w$, i.e., $w[i,j] = a_i\dots a_j$.
We write $u \sqsubseteq w$ if $w = uv$ for $u,v \in \Sigma^*$.
A subset $L \subseteq \Sigma^*$ is called \textit{language} over $\Sigma$.

A \textit{ranked alphabet} $\Sigma$ is an alphabet where each letter $f \in \Sigma$ has a finite set of arities $rk(f) \subseteq \mathbbm{N}$.
The set of letters of arity $i$ is denoted by $\Sigma_i$.
A tree domain $\dom{}$ is a non-empty finite subset of $(\mathbbm N\setminus\{0\})^*$ such that $\dom{}$ is prefix-closed and for each $u \in (\mathbbm N\setminus\{0\})^*$ and $i \in \mathbbm N\setminus\{0\}$ if $ui \in \dom{}$, then $uj \in \dom{}$ for all $1 \leq j < i$.
We speak of $ui$ as successor of $u$ for each $u \in \dom{}$ and $i \in \mathbbm N\setminus\{0\}$. 

A (finite $\Sigma$-labeled) \textit{tree} is a pair $t = (\dom{t},\val{t})$ with a mapping $\val{t}: \dom{t} \rightarrow \Sigma$ such that for each node $u \in \dom{t}$ the number of successors of $u$ corresponds to a rank of $\val{t}(u)$.
The set of all $\Sigma$-labeled trees is denoted by $T_{\Sigma}$.
A subset $T \subseteq T_{\Sigma}$ is called \textit{tree language} over $\Sigma$.

A \textit{subtree} $t|_u$ of a tree $t$ at node $u$ is defined by $\dom{t|_u} = \{ v \in \mathbbm{N}^* \mid uv \in \dom{t} \}$ and $\val{t|_u}(v) = \val{t}(uv)$ for all $v \in \dom{t|_u}$.
In order to formalize concatenation of trees, we introduce the notion of special trees.
A \textit{special tree} over $\Sigma$ is a tree over \(\Sigma \cupcdot \{\circ\}\) such that $\circ$ occurs exactly once at a leaf.
Given $t \in T_{\Sigma}$ and $u \in \dom{t}$, we write $t[\circ/u]$ for the special tree that is obtained by deleting the subtree at u and replacing it by $\circ$.
Let $S_{\Sigma}$ be the set of special trees over $\Sigma$.
For $t \in T_{\Sigma}$ or $t \in S_{\Sigma}$ and $s \in S_{\Sigma}$ let the \textit{concatenation} $t \cdot s$ be the tree that is obtained from $t$ by replacing $\circ$ with $s$.

Let $X_n$ be a set of $n$ variables $\{x_1,\dots,x_n\}$ and $\Sigma$ be a ranked alphabet.
We denote by $T_{\Sigma}(X_n)$ the set of all trees over $\Sigma$ which additionally can have variables from $X_n$ at their leaves.
Let $X = \bigcup_{n > 0} X_n$.
For $t \in T_{\Sigma}(X_n)$ let $t[x_1 \leftarrow t_1, \dots, x_n \leftarrow t_n]$ be the tree that is obtained by substituting each occurrence of $x_i \in X_n$ by $t_i \in T_{\Sigma}(X)$ for every $1 \leq i \leq n$.

A tree from $T_{\Sigma}(X_n)$ such that all variables from $X_n$ occur exactly once and in the order $x_1,\dots,x_n$ when reading the leaf nodes from left to right, is called $n$-\textit{context} over $\Sigma$.
A special tree can be seen as an $1$-context.
If $C$ is an $n$-context and $t_1, \dots, t_n \in T_{\Sigma}(X)$ we write $C[t_1,\dots,t_n]$ instead of $C[x_1 \leftarrow t_1, \dots, x_n \leftarrow t_n]$.

\medskip
\noindent \textbf{Tree automata.}
Tree automata can be viewed as a straightforward generalization of finite automata on finite words, when words are interpreted as trees over unary symbols.
For a detailed introduction to tree automata see e.g. \cite{gesceg1984tree} or \cite{tata2007}.

Let $\Sigma = \bigcup_{i = 1}^m \Sigma_i$ be a ranked alphabet. A \textit{non-deterministic top-down tree automaton} (an N$\downarrow$TA) over $\Sigma$ is of the form $\mathcal A = (Q, \Sigma, Q_0, \Delta)$ consisting of a finite set of states $Q$, a set $Q_0 \subseteq Q$ of initial states, and $\Delta \subseteq \bigcup_{i=0}^m (Q \times \Sigma_i \times Q^i)$ is the transition relation.
For $i = 0$, we identify $Q \times \Sigma_i \times Q^i$ with $Q \times \Sigma_0$.

Let $t$ be a tree and $\mathcal A$ be an N$\downarrow$TA, a \textit{run} of $\mathcal A$ on $t$ is a mapping $\rho: \dom{t} \rightarrow Q$ compatible with $\Delta$, i.e., $\rho(\varepsilon) \in Q_0$ and for each node $u \in \dom{t}$, if $\val{t}(u) \in \Sigma_i$ with $i \geq 0$, then $(\rho(u),\val{t}(u),\rho(u1),\dots,\rho(ui)) \in \Delta$.
A tree $t \in T_{\Sigma}$ is \textit{accepted} if, and only if, there is a run of $\mathcal A$ on $t$.
The tree language \textit{recognized} by $\mathcal A$ is $T(\mathcal A) = \{ t \in T_{\Sigma} \mid \mathcal A \text{ accepts } t\}$. 

A tree language $T \subseteq T_{\Sigma}$ is called \textit{regular} if $T$ is recognizable by a non-deterministic top-down tree automaton.
As the class of regular word languages, the class of regular tree languages is closed under Boolean operations.

A top-down tree automaton $\mathcal A = (Q,\Sigma,Q_0,\Delta)$ is \textit{deterministic} (a D$\downarrow$TA) if the set $Q_0$ is a singleton set and for each $f \in \Sigma_i$ and each $q \in Q$ there is at most one transition $(q,f,q_1,\dots,q_i) \in \Delta$.
However, non-deterministic and deterministic top-down automata are not equally expressive.

An extension to regular tree languages are (binary) \textit{tree-automatic relations}.
A way for a tree automaton to read a tuple of finite trees is to use a ranked vector alphabet. Thereby, all trees are read in parallel, processing one node from each tree in a computation step.
Hence, the trees are required to have the same domain.
Therefore we use a padding symbol to extend the trees if necessary.
Formally, this is done in the following way.

Let $\Sigma$, $\Gamma$ be ranked alphabets and let $\Sigma_{\bot} = \Sigma \cupcdot \{\bot\}$, $\Gamma_{\bot} = \Gamma \cupcdot \{\bot\}$.
The \textit{convolution} of $(t_1,t_2)$ with $t_1 \in T_{\Sigma}$, $t_2 \in T_{\Gamma}$ is the $\Sigma_{\bot} \times \Gamma_{\bot}$-labeled tree $t = t_1 \otimes t_2$ defined by $\dom{t} = \dom{t_1} \cup \dom{t_2}$, and $\val{t}(u) = (\val{t_1}^{\bot}(u), \val{t_2}^{\bot}(u))$ with $rk(\val{t}(u)) = max\left\{rk(\val{t_1}^{\bot}(u)), rk(\val{t_2}^{\bot}(u))\right\}$ for all $u \in \dom{t}$, where $\val{t_i}^{\bot}(u) = \val{t_i}(u)$ if $u \in \dom{t_i}$ and $\val{t_i}^{\bot}(u) = \bot$ otherwise for $i \in \{1,2\}$.
As a special case, given $t \in T_{\Sigma}$, we define $t \otimes \bot$ to be the tree with $\dom{t \otimes \bot} = \dom{t}$ and $\val{t \otimes \bot}(u) = (\val{t}(u),\bot)$ for all $u \in \dom{t}$.
Analogously, we define $\bot \otimes t$.
We define the \textit{convolution of a tree relation} $R \subseteq T_{\Sigma} \times T_{\Gamma}$ to be the tree language $T_R := \{ t_1 \otimes t_2 \mid (t_1,t_2) \in R\}$.

We call a (binary) relation $R$ \textit{tree-automatic} if there exists a regular tree language $T$ such that $T = T_R$.
For ease of presentation, we say a tree automaton $\mathcal A$ recognizes $R$ if it recognizes the convolution $T_R$ and denote by $R(\mathcal A)$ the induced relation $R$.

A \textit{uniformization} of a relation $R \subseteq X \times Y$ is a function $f_R: X \to Y$ such that for each domain element $x$ the pair $(x,f_R(x))$ is in the relation, i.e., $(x,f_R(x)) \in R \text{ for all } x \in dom(R)$.
In the following, we are interested for a given tree-automatic relation, whether there exists a uniformization which can be realized by a tree transducer.

\medskip
\noindent \textbf{Tree Transducers.}
Tree transducers are a generalization of word transducers.
As top-down tree automata, a top-down tree transducer reads the tree from the root to the leafs, but can additionally in each computation step produce finite output trees which are attached to the already produced output.
For an introduction to tree transducers the reader is referred to \cite{tata2007}.

A \textit{top-down tree transducer} (a TDT) is of the form $\mathcal T = (Q,\Sigma,\Gamma,q_0,\Delta)$ consisting of a finite set of states $Q$, a finite input alphabet $\Sigma$, a finite output alphabet $\Gamma$, an initial state $q_0 \in Q$, and $\Delta$ is a finite set of transition rules of the form
\begin{center}
 $q(f(x_1,\dots,x_i)) \rightarrow w[q_1(x_{j_1}),\dots,q_n(x_{j_n})]$,
\end{center}
where $f \in \Sigma_i$, $w$ is an $n$-context over $\Gamma$, $q,q_1,\dots,q_n \in Q$ and $j_1,\dots,j_n \in \{1,\dots,i\}$, or
\begin{center}
 $q(x_1) \rightarrow w[q_1(x_1),\dots,q_n(x_1)]$\quad ($\varepsilon$-transition),
\end{center}
where $f \in \Sigma_i$, $w$ is an $n$-context over $\Gamma$, and $q,q_1,\dots,q_n \in Q$.
A top-down tree transducer is \textit{deterministic} (a DTDT) if it contains no $\varepsilon$-transitions and there are no two rules with the same left-hand side.

A \textit{configuration} of a top-down tree transducer is a triple $c = (t,t',\varphi)$ of an input tree $t \in T_{\Sigma}$, an output tree $t' \in T_{\Gamma \cup Q}$ and a function $\varphi: D_{t'} \rightarrow \dom{t}$, where
\begin{itemize}
 \item $\val{t'}(u) \in \Gamma_i$ for each $u \in \dom{t'}$ with $i > 0$ successors
 \item $\val{t'}(u) \in \Gamma_0$ or $\val{t'}(u) \in Q$ for each leaf $u \in \dom{t'}$ 
 \item $D_{t'} \subseteq \dom{t'}$ with $D_{t'} = \{ u \in \dom{t'} \mid \val{t'}(u) \in Q\}$ \\ \quad \phantom{\ } \hfill \normalsize{($\varphi$ maps every node from the output tree $t'$ that has a state-label to a node of the input tree $t$)}
\end{itemize}
Let $c_1 = (t,t_1,\varphi_1), c_2 = (t,t_2,\varphi_2)$ be configurations of a top-down tree transducer.
We define a successor relation $\rightarrow_{\mathcal T}$ on configurations as usual by applying one rule, which looks formally (for the application of a non-$\varepsilon$-rule) as follows:
\begin{center}
\parbox{\textwidth}{
$c_1 \rightarrow_{\mathcal T}c_2 :\Leftrightarrow \left \{
\begin{array}{l}
  \exists u \thinspace \in \textnormal{dom}_{t_1}, q \in Q, v \in \textnormal{dom}_t \textnormal{ with } \textnormal{val}_{t_1}(u) = q \textnormal{ and } \varphi_1(u) = v\\
  \exists \thinspace q(\textnormal{val}_t(v)(x_1,\dots,x_i)) \rightarrow w[q_1(x_{j_1}),\dots,q_n(x_{j_n})] \in \Delta\\
  t_2 = s \cdot w[q_1,\dots,q_n] \textnormal{ with } s = t_1[\circ/u]\\
  \varphi_2 \textnormal{ with } D_{t_2} = D_{t_1}\setminus\{u\} \cup \{u_i \mid u \sqsubseteq u_i, \textnormal{val}_{t_2}(u_i) = q_i, 1 \leq i \leq n\}\\
  \forall \thinspace u' \in D_{t_1}\setminus \{u\}: \varphi_2(u') = \varphi_1(u')\\
  \forall \thinspace u_i, u \sqsubseteq u_i, \textnormal{val}_{t_2}(u_i) = q_i: \varphi_2(u_i) = v.j_i
\end{array}\right .$
}
\end{center}
Furthermore, let $\rightarrow_{\mathcal T}^*$ be the reflexive and transitive closure of $\rightarrow_{\mathcal T}$ and  $\rightarrow_{\mathcal T}^n$ the reachability relation for $\rightarrow_{\mathcal T}$ in $n$ steps.
The relation $R(\mathcal T) \subseteq T_{\Sigma} \times T_{\Gamma}$ induced by a top-down tree transducer $\mathcal T$ is $$R(\mathcal T) = \{(t,t') \mid (t,q_0,\varphi_0) \rightarrow_{\mathcal T}^* (t,t',\varphi') \text{ with } \varphi_0(\varepsilon)=\varepsilon, \varphi' \text{ is empty} \text{ and } t' \in T_{\Gamma}\}.$$
For a tree $t \in T_{\Sigma}$ let $\mathcal T(t)$ be the final transformed output of $\mathcal T$ for $t$.
The class of relations definable by TDTs is called the class of \textit{top-down tree transformations}.

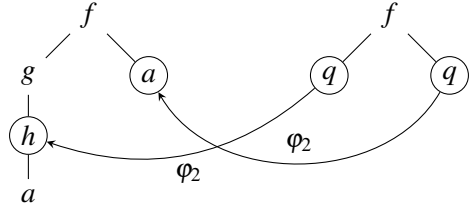
\begin{figure}[t!]
\centering
\begin{tikzpicture}
  \tikzstyle{level 1}=[sibling distance=16mm]
  \tikzstyle{level 2}=[sibling distance=8mm]
  \tikzstyle{level 3}=[sibling distance=8mm]
  \tikzstyle{level 4}=[sibling distance=8mm]
  \begin{scope}[xshift = -4cm]
  \path[level distance=8mm] node (root){$f$}
    child{
      node(0){$g$}
	child{
	  node(00)[draw = black, circle, inner sep=0pt, minimum size = 5mm]{$h$}
	    child{
	      node(000){$a$}
	    }
	}
    }
    child{
      node(1)[draw = black, circle, inner sep=0pt, minimum size = 5mm]{$a$}
    }
  ;
  \end{scope}
  \path[level distance=8mm] node (root){$f$}
    child{
      node(a)[draw = black, circle, inner sep=0pt, minimum size = 5mm]{$q$}
    }
    child{
      node(b)[draw = black, circle, inner sep=0pt, minimum size = 5mm]{$q$}
    }
    (a) edge [bend left, ->, >=stealth, below, color = black] node {\small{$\varphi_2$}} (00)
    (b) edge [bend left = 60, ->, >=stealth, above, color = black] node {\small{$\varphi_2$}} (1)
  ;
\end{tikzpicture}
\caption{The configuration $c_2 = (t,f(q,q),\varphi_2)$ of $\mathcal T$ on $t$ from Example \ref{ex:transducer}.}
\label{fig:configuration}
\end{figure}

\begin{example}\label{ex:transducer}
 Let $\Sigma$ be a ranked alphabet given by $\Sigma_2 = \{f\}$, $\Sigma_1 = \{g,h\}$, and $\Sigma_0 = \{a\}$.
 Consider the TDT $\mathcal T$ given by $(\{q\},\Sigma,\Sigma,\{q\},\Delta)$ with $\Delta$ $=$ $\{$ $q(a) \rightarrow a$, $q(g(x_1)) \rightarrow q(x_1)$, $q(h(x_1)) \rightarrow h(q(x_1))$, $q(f(x_1,x_2)) \rightarrow f(q(x_1),q(x_2))$  $\}$.
 For each $t \in T_{\Sigma}$ the transducer deletes all occurrences of $g$ in $t$.
 
 Consider $t := f(g(h(a)),a)$. A possible sequence of configurations of $\mathcal T$ on $t$ is $c_0 \rightarrow_{\mathcal T}^5 c_5$ such that $c_0 := (t,q,\varphi_0)$ with $\varphi_0(\varepsilon) = \varepsilon$, $c_1 := (t,f(q,q),\varphi_1)$ with $\varphi_1(1) = 1$, $\varphi_1(2) = 2$, $c_2 := (t,f(q,q),\varphi_2)$ with $\varphi_2(1) = 11$, $\varphi_2(2) = 2$, $c_3 := (t,f(q,a),\varphi_3)$ with $\varphi_3(1) = 11$, $c_4 := (t,f(h(q),a),\varphi_4)$ with $\varphi_4(11) = 111$, and $c_5 := (t,f(h(a),a),\varphi_5)$.
 A visualization of $c_2$ is shown in Figure \ref{fig:configuration}.
\end{example}

\medskip
\noindent \textbf{Games.}
A safety game $\mathcal G = (V, V_0,V_1,E,S)$ is played by two players, Player 0 and Player 1, on a directed game graph $G = (V,E)$, where
\begin{itemize}
 \item $V = V_0 \cupcdot V_1$ is a partition of the vertices into positions $V_0$ belonging to Player 0 and positions $V_1$ belonging to Player 1,
 \item $E \subseteq V \times V$ is the set of allowed moves, and
 \item $S \subseteq V$ is a set of safe vertices.
\end{itemize}
A play is a maximal finite or infinite sequence $v_0v_1v_2\dots$ of vertices compatible to the edges of the game graph starting from an initial vertex $v_0 \in V$.
A play is maximal if it is either infinite or it ends in a vertex without outgoing edges.
Player 0 wins a play if it stays inside the safe region, i.e., $v_i \in S$ for all $i$.

Let $i \in \{0,1\}$, a strategy for Player $i$ is a function $\sigma_i: V^*V_i \rightarrow V$ such that $\sigma_i(v_0\dots v_n) = v_{n+1}$ implies that $(v_n,v_{n+1}) \in E$.
A strategy $\sigma_i$ is a winning strategy from a vertex $v_0 \in V$ for Player $i$ if the player wins every play starting in $v_0$, no matter how the opponent plays, if Player $i$ plays according $\sigma_i$.

Safety games are positionally determined, cf.\ \cite{GraedelThoWil02}, i.e., for each vertex $v \in V$ one of the players has a winning strategy from $v$.
Furthermore, the player always has a positional winning strategy $\sigma$, meaning that the strategy does not consider the previously seen vertices, but only the current vertex.
More formally, a positional strategy $\sigma_i$ for Player $i$ is a mapping $\sigma_i: V_i \rightarrow V$ such that $(v,\sigma_i(v)) \in E$ for all $v \in V_i$.

\section{Bounded Delay}
\label{sec:3}
In this section we investigate uniformization of tree-automatic relations in the class of top-down tree transformations.
We restrict ourselves in the scope of this section to D$\downarrow$TA-recognizable relations with D$\downarrow$TA-recognizable domain.
For each valid input tree the transducer selects one output tree, on each other input tree which is not part of the domain the transducer may behave arbitrarily.
To distinguish between these issues, we will speak of uniformization with input validation and uniformization without input validation.
For ease of presentation, in this and the following section we will only consider relations with total domain.
Later on, we will briefly describe how we can deal with relations whose domain is not total but deterministic top-down tree automatic.

For the remainder of this paper, let $R \subseteq T_{\Sigma} \times T_{\Gamma}$ be a deterministic top-down tree automaton-definable relation with total domain and let $\mathcal A = (Q_{\mathcal A},\Sigmabot \times \Gammabot,q_0^{\mathcal A},\Delta_{\mathcal A})$ be a D$\downarrow$TA that recognizes $R$.
For $q \in Q_{\mathcal A}$, let $\mathcal A_q$ be the automaton that results from $\mathcal A$ by using $q$ as single initial state.

To begin with, we investigate the connection between input and output.
A TDT $\mathcal T$ possibly reaches a point such that the position $v$ of the input symbol under consideration and the position $u$ of the correspondingly produced output are different (formally, the mapping $\varphi$ from the configuration maps $u$ to a node $v \neq u$).
As a consequence, if $u$ and $v$ lie on divergent paths, then $\mathcal T$ selects the output at $u$ (and below $u$) independent of the subtree at $u$ of the input tree.
That means, in order to satisfy a given specification, if $\mathcal T$ selects the output independent of the input from some point on, then the produced output tree must match all possible input trees from then on.
Such cases are considered in the Lemma below.

\begin{lemma}\label{lemma:edgecontraints}
 Let $q \in Q_{\mathcal A}$.
 It is decidable whether the following holds:
 \begin{enumerate}
  \item $\forall t \in T_{\Sigma}: t \otimes \bot \in T(\mathcal A_q)$,
  \item $\exists t' \in T_{\Gamma}: \bot \otimes t' \in T(\mathcal A_q)$,
  \item $\exists t' \in T_{\Gamma}\thinspace \forall t \in T_{\Sigma}: t \otimes t' \in T(\mathcal A_q)$.
 \end{enumerate}
\end{lemma}

Now, we formally define what is meant by output delay.
Let $\mathcal T$ be a TDT and let $c = (t,t',\varphi)$ be a configuration of $\mathcal T$.
Consider a node $u \in D_{t'}$.
If $|\varphi(u)| \geq |u|$, we say the transducer has an \textit{output delay} of $|\varphi(u)| - |u|$.
If there is a $k \in \mathbbm N$ such that for all reachable configurations $c = (t,t',\varphi)$ of $\mathcal T$ holds that for all $u \in D_{t'}$ the output delay $|\varphi(u)| - |u|$ is at most $k$, then the output delay is bounded to $k$.

We will solve the following problem.

\begin{theorem}\label{theorem:BoundedDelayDecidableTDT}
 Given $k > 0$, it is decidable whether a given D$\downarrow$TA-recognizable relation with total domain has a uniformization by a deterministic top-down tree transducer with output delay bounded to $k$.
\end{theorem}

Before we present a decision procedure, we introduce some notations that will simplify the presentation. 
Given $\Sigma = \bigcup_{i=0}^m \Sigma_i$, let $\mathrm{dir}_\Sigma = \{1,\dots,m\}$ be the set of directions compatible with $\Sigma$.
For $\Sigma = \bigcup_{i=0}^m \Sigma_i$, the set $\mathrm{Path}_{\Sigma}$ of labeled paths over $\Sigma$ is defined inductively by:
\begin{itemize}
 \item $\varepsilon$ is a labeled input path and each $f \in \Sigma$ is a labeled input path,
 \item given a labeled input path $\pi = x \cdot f$ with $f \in \Sigma_i\thinspace (i > 0)$ over $\Sigma$, then $\pi \cdot jg$ with $j \in \{1,\dots,i\}$ and $g \in \Sigma$ is a labeled input path.
\end{itemize}
For $\pi \in \mathrm{Path}_{\Sigma}$, we define the path $path$ and the word $lbls$ induced by $\pi$ inductively by:
\begin{itemize}
 \item if $\pi = \varepsilon$ or $\pi = f$, then $path(\varepsilon) = path(f) = \varepsilon$, $lbls(\varepsilon)$ = $\varepsilon$ and $lbls(f) = f$,
 \item if $\pi = x \cdot jf$ with $j \in \mathbbm{N},\thinspace f \in \Sigma$, then $path(\pi) = path(x)\cdot j$, $lbls(\pi) = lbls(x) \cdot f$.
\end{itemize}
The length $||\thinspace||$ of a labeled path over $\Sigma$ is the length of the word induced by its path, i.e., $||\pi|| = |lbls(\pi)|$.

For $\pi \in \mathrm{Path}_{\Sigma}$ with $||\pi|| = k$ let
\begingroup
\setlength{\abovedisplayskip}{.5\columnsep }
\setlength{\belowdisplayskip}{.5\columnsep }
\begin{equation*}
 T_{\Sigma}^\pi := \{ t \in T_{\Sigma} \mid \val{t}\bigl(path(\pi)[1\dots(i-1)]\bigr) = lbls(\pi)[i] \text{ for } 1 \leq i \leq k\}
\end{equation*}
\endgroup
be the set of trees $t$ over $\Sigma$ such that $\pi$ is a prefix of a labeled path through $t$.
For $\Pi \subseteq \mathrm{Path}_{\Sigma}$ let
\begingroup
\setlength{\abovedisplayskip}{.5\columnsep }
\setlength{\belowdisplayskip}{.5\columnsep }
\begin{equation*}
 T_{\Pi} := \{ t \in T_{\Sigma} \mid \exists \pi \in \Pi \text{ and } t \in T_{\Sigma}^{\pi}\}
\end{equation*}
\endgroup
be the set of trees such that each tree contains a labeled path starting with $\pi$ for some $\pi \in \Pi$.

For $t \in T_{\Sigma}$ and $u \in \mathrm{dir}_\Sigma^*$, let $||t||^u := max\{ |v| \mid v \in \dom{t} \text{ and } (u \sqsubseteq v \text{ or } v \sqsubseteq u)\}$ be the length of a maximal path through $t$ along $u$.

Now, in order to solve the above decision problem, we consider a safety game between two players.
The procedure is similar to a decision procedure presented in \cite{uniformizationAutomataTheory}, where the question whether a uniformization of an automatic word relation by a word transducer exists, is reduced to the existence of winning strategies in a safety game.
The game is played between \In\ and \Out, where \In\ can follow any path from the root to a leaf in an input tree such that \In\ plays one input symbol at a time.
\Out\ can either react with an output symbol, or delay the output and react with a direction in which \In\ should continue with his input sequence.

The vertices in the game graph keep track of the state of $\mathcal A$ on the input combined with the output on the same path and additionally of the input that is ahead, which is bounded to $k$.
We will see that it is not necessary to consider situations where input and output are on divergent paths.
The intuition behind this is that D$\downarrow$TAs cannot compare information on divergent paths through an input tree.
Formally, the game graph $G_{\mathcal A}^k$ is constructed as follows.
\begin{itemize}
 \item $\Vin = \{ \bigl(q,\pi j\bigr) \in Q_{\mathcal A} \times \mathrm{Path}_\Sigma \cdot \mathrm{dir}_\Sigma \mid \|\pi\| \leq k, \pi \in \mathrm{Path}_\Sigma, j \in \mathrm{dir}_\Sigma\} \cup 2^{Q_{\mathcal A}}$ is the set of vertices of player \In.
 \item $\Vout = \{ \bigl(q,\pi\bigr) \in Q_{\mathcal A} \times \mathrm{Path}_\Sigma  \mid \|\pi\| \leq k\}$ is the set of vertices of player \Out.
 \item From a vertex of \In\ the following moves are possible:
  \begin{enumerate}[i)]
   \item $\bigl(q,\pi j\bigr) \rightarrow \bigl(q,\pi jf\bigr)$ for each $f \in \Sigma$ if $\|\pi\| < k$ \hfill (delay; \In\ chooses the next input)
   \item $ \{q_1,\dots,q_n\} \rightarrow \bigl(q_i,f\bigr)$ for each $i \in \{1,\dots,n\}$
         \begin{flushright}(no delay; \In\ chooses the next direction and input)
         \end{flushright}
  \end{enumerate}
 \item From a vertex of \Out\ the following moves are possible:
  \begin{enumerate}[i)]\setcounter{enumi}{2}
   \item $\bigl(q,f\bigr) \overset{r}{\rightarrow} \{q_1,\dots,q_i\}$ if there is $r = (q,(f,g),q_1,\dots,q_n) \in \Delta_{\mathcal A}$, $f \in \Sigma$ is $i$-ary, $g \in \Sigmabot$ is $j$-ary, and if $j > i$, there exist trees $t_{i+1},\dots,t_j \in T_{\Gamma}$ such that $\bot \otimes t_l \in T(\mathcal A_{q_l})$ for all $i < l \leq j$.
      \begin{flushright}
  (no delay; \Out\ applies a transition; \Out\ can pick output trees for all directions where the input has ended; \In\ can continue from the other directions)      
      \end{flushright}
   \item $\bigl(q,\pi j'f'\bigr) \overset{r}{\rightarrow} \bigl(q',\pi'j'f'\bigr)$ for each $g \in \Gamma_{\bot}$ such that $\pi = fj\pi'$, there is $r = (q,(f,g),q_1,\dots,q_n) \in \Delta_{\mathcal A}$ with $q' = q_j$, and for each $l \neq j$ with $l \in \{1,\dots,n\}$ holds
    \begin{itemize}
     \item if $l \leq rk(f),rk(g)$, then $\exists t' \in T_{\Gamma} \forall t \in T_{\Sigma}: t \otimes t' \in T(\mathcal A_{q_l})$
     \item if $rk(g) < l \leq rk(f)$, then $\forall t\in T_{\Sigma}: t \otimes \bot \in T(\mathcal A_{q_l})$
     \item if $rk(f) < l \leq rk(g)$, then $\exists t' \in T_{\Gamma}: \bot \otimes t' \in T(\mathcal A_{q_l})$
    \end{itemize}
      \begin{flushright}
  (delay; \Out\ applies a transition, removes the leftmost input and advances in direction of the labeled path ahead; \Out\ can pick output trees for all divergent directions)  
      \end{flushright}
   \item $\bigl(q,\pi jf\bigr) \rightarrow \bigl(q,\pi jf j'\bigr)$ for each $j' \in \{1,\dots,i\}$ for $f \in \Sigma_i$ if $\|\pi jf\| < k$
    \begin{flushright}
  (\Out\ delays and chooses a direction from where \In\ should continue)   
      \end{flushright}
  \end{enumerate}
 \item The initial vertex is $\{q_0^{\mathcal A}\}$.
\end{itemize}
Note that the game graph can effectively be constructed, because Lemma \ref{lemma:edgecontraints} implies that it is decidable whether the edge constraints are satisfied.

The winning condition should express that player \Out\ loses the game if the input can be extended, but no valid output can be produced.
This is represented in the game graph by a set of bad vertices $B$ that contains all vertices of \Out\ with no outgoing edges.
If one of these vertices is reached during a play, \Out\ loses the game.
Thus, we define $\mathcal G_{\mathcal A}^k = (G_{\mathcal A}^k, V \setminus B)$ as safety game for \Out.

\begin{example}\label{ex:simpleCase}
 Let $\Sigma$ be an input alphabet given by $\Sigma_2 = \{f\}$ and $\Sigma_0 = \{a\}$ and let $\Gamma$ be an output alphabet given by $\Gamma_2 = \{f,g\}$ and $\Gamma_0 = \{b\}$.
 Consider the relation $R$ that contains exactly the pairs of trees $(t,t') \in T_{\Sigma} \times T_{\Gamma}$ such that $t$ and $t'$ have the same domain and on every path through $t'$ occurs an $f$ if $t \neq a$.

 It is easy to see that D$\downarrow$TA $\mathcal A = (\{q_0,q,q_f\},\Sigma \times \Gamma,q_0,\Delta_{\mathcal A})$ with $\Delta_{\mathcal A}$ $=$ $\{(q_0,(a,b))$, $(q_0,(f,f),q_f,q_f)$, $(q_0,(f,g),q,q)$, $(q,(f,f),q_f,q_f)$, $(q,(f,g),q,q)$, $(q_f,(a,b))$, $(q_f,(f,f),q_f,q_f)$, $(q_f,(f,g),q_f,q_f)\}$ recognizes $R$.
 For $k = 1$, the corresponding game graph $G_{\mathcal A}^1$ is depicted in Figure \ref{fig:simpleCase}.
\end{example}

\begin{figure}[t!]
\centering
\begin{tikzpicture}[scale=0.9]

\tikzstyle{playerIn}=[rectangle, draw, inner sep=1pt,minimum size = 6mm, color=black]
\tikzstyle{playerOut}=[rectangle, rounded corners, draw, inner sep=1pt, minimum size = 6mm, color=black]
\tikzstyle{edge}=[->,>=stealth]
\tikzstyle{thickedge}=[->,>=stealth,very thick]

\node (99)	[draw=none]	at (0,7){};
\node (0) 	[playerIn] 	at (0, 6) 	{$\{q_0\}$};
\node (1)	[playerOut] 	at (-2, 5)	{$q_0,a$};
\node (2)	[playerOut]	at (2, 5)	{$q_0,f$};
\node (3)	[playerIn] 	at (-4, 3.5)	{$\emptyset$};
\node (4)	[playerIn] 	at (0, 3.5)	{$\{q_f\}$};
\node (5)	[playerIn] 	at (4, 3.5)	{$\{q\}$};
\node (6)	[playerOut] 	at (-2, 2)	{$q_f,a$};
\node (7)	[playerOut] 	at (0, 2)	{$q_f,f$};
\node (8)	[playerOut] 	at (6, 2)	{$q,a$};
\node (9)	[playerOut] 	at (4, 2)	{$q,f$};

\draw [edge] (99) to (0);

\draw [edge] (0) to (1);
\draw [edge] (0) to (2);

\draw [thickedge] (1) to node[above left]{$b$}(3);
\draw [thickedge] (2) to node[above left]{$f$}(4);
\draw [edge] (2) to node[above right]{$g$}(5);

\draw [thickedge] (6) to node[above right]{$b$}(3);
\draw [edge] (4) to (6);
\draw [edge] (4) to [bend right = 15] (7);
\draw [thickedge] (7) to [bend right = 15] node[midway,right]{$f,g$} (4);
\draw [edge] (5) to (8);
\draw [edge] (5) to [bend right = 15] (9);
\draw [edge] (9) to [bend right = 15] node[midway,right]{$g$} (5);
\draw [edge] (9) to node[midway,above]{$f$} (4);

\end{tikzpicture}
\caption{The game graph $G_{\mathcal A}^1$ constructed from the D$\downarrow$TA $\mathcal A$ from Example \ref{ex:simpleCase}.
A possible winning strategy for \Out\ in $\mathcal G_{\mathcal A}^1$ is emphasized in the graph.}
\label{fig:simpleCase}
\end{figure}

The following two lemmata show that from the existence of a winning strategy a top-down tree transducer that uniformizes the relation can be obtained and vice versa.

\begin{lemma}\label{lemma:StrategytoBoundedDelayTDT}
If \Out\ has a winning strategy in $\mathcal G_{\mathcal A}^k$, then $R$ has a uniformization by a DTDT in which the output delay is bounded to $k$.
\end{lemma}

\ifproofs
\begin{proof}
 Assume that \Out\ has a winning strategy in the safety game $\mathcal G_{\mathcal A}^k$, then there is also a positional one.
 We can represent a positional winning strategy by a function $\sigma: \Vout \rightarrow \Delta_{\mathcal A} \cup \mathrm{dir}_{\Sigma}$, because \Out\ either plays one output symbol (corresponding to a unique transition in $\Delta_{\mathcal A}$), or a new direction for an additional input symbol.

 We construct a deterministic TDT $\mathcal{T} = (Q_{\mathcal A} \cup \{\bigl(q,\pi j\bigr) \mid \bigl(q,\pi jf\bigr) \in \Vout\},\Sigma,\Gamma,q_0^{\mathcal A},\Delta)$ from such a positional winning strategy $\sigma$ as follows:
 \begin{enumerate}[a)]
  \item For each $\sigma: \bigl(q,f\bigr) \overset{r}{\mapsto} \{q_1,\dots,q_i\}$ with $r = (q,(f,g),q_1,\dots,q_n) \in \Delta_{\mathcal A}$:
   \begin{itemize}
    \item add $q(f(x_1,\dots,x_i)) \rightarrow g(q_1(x_1),\dots,q_j(x_j))$ to $\Delta$ if $j\leq i$, or
    \item add $q(f(x_1,\dots,x_i)) \rightarrow g(q_1(x_1),\dots,q_i(x_i),t_{i+1},\dots,t_{j})$ to $\Delta$ if $j > i$
   \end{itemize}
   where $f \in \Sigma_i$, $g \in \Gamma_ j$ and $t_{i+1},\dots,t_{j} \in T_{\Gamma}$ chosen according to the $r$-edge constraints in $\bigl(q,f\bigr)$.
  \item \mbox{For each $\sigma: \bigl(q,\pi jf\bigr) \mapsto \bigr(q,\pi jfj'\bigr)$ add $\bigl ( q,\pi j \bigr )(f(x_1,\dots,x_i))$} $\rightarrow \bigl( q,\pi jfj' \bigr )(x_{j'})$ to $\Delta$.
 \end{enumerate}
 If the strategy $\sigma$ defines a sequence of moves of \Out\ inside vertices of \Vout, that is a sequence of moves of type iv), then this corresponds to an output sequence that is produced without reading further input.
 Each output of these moves can be represented by a special tree $s$ as follows.
 A move of type iv) has the form $(q,fj\pi) \overset{r}{\rightarrow} (q',\pi)$ with $r = (q,(f,g),q_1,\dots,q_n)$ and $q' = q_j$.
 Then, let $s = g(t_1,\dots,t_{j-1},\circ,t_{j+1},\dots,t_n) \in S_{\Sigma}$ be the special tree, where each $t_l \in T_{\Gamma}$ is choosen according to the $r$-edge constraints in $(q,fj\pi)$ for $l \neq j, 1 \leq l \leq n$.
 Eventually, the strategy defines a move of \Out\ of type iii) or v) to a node of \Vin, otherwise $\sigma$ is not a winning strategy.
 These parts of the strategy are transformed as follows:
  \begin{enumerate}[a)]
  \setcounter{enumi}{2}
   \item
    For each $\bigl(q,\pi
 jf\bigr) \overset{r_1}{\rightarrow} \dots \overset{r_{l-1}}{\rightarrow} \bigl(q',\pi'jf\bigr) \overset{}
 {\rightarrow} \bigl(q',\pi'jfj'\bigr) $ add \\ $\bigl ( q,\pi j \bigr )(f(x_1,\dots,x_i)) \rightarrow s_1 \cdot \ldots \cdot s_{l-1} \cdot \bigl( q',\pi'jfj' \bigr )(x_{j'})$ to $\Delta$, where each $s_i \in S_{\Gamma}$ is a special tree corresponding to the $r_i$-edge in the $i$th move.
   \item
    For each $\bigl(q,\pi jf\bigr) \overset{r_1}{\rightarrow} \dots \overset{r_{l-1}}{\rightarrow} \bigl(q',f\bigr) \overset{r_l} {\rightarrow} \{q_1,\dots,q_i\}$ add $\bigl ( q,\pi j \bigr )(f(x_1,\dots,x_i)) \rightarrow s_1 \cdot \ldots \cdot s_{l-1} \cdot s$ to $\Delta$, where each $s_i \in S_{\Gamma}$ is a special tree corresponding to the $r_i$-edge in the $i$th move and $s$ is an output corresponding to $r_l$ constructed as described in step a).
  \end{enumerate}
 We now verify that $\mathcal T$ defines a uniformization of $R$.
 Let $t \in T_{\Sigma}$.
 We can show by induction on the number of steps needed to reach a configuration from the initial configuration $(t,q_0^{\mathcal A},\varphi_0)$ that for each configuration $c = (t,t',\varphi)$ such that $D_{t'} \neq \emptyset$, in other words $t' \notin T_{\Gamma}$, there exists a successor configuration $c'$.
 Thus, $(t,\mathcal T(t)) \in R$.
 
 In $\mathcal T$ the output delay is bounded to $k$, because the existence of a winning strategy $\sigma$ guarantees that from a vertex $(q,\pi)$ with $|\pi| = k$, that is reachable by playing according to $\sigma$, a move of \Out \ follows.
 It follows from the construction that $\mathcal T$ produces output accordingly.
\end{proof}
\fi

The size of $G_{\mathcal A}^k$ is at most $Q_{\mathcal A} \cdot (|\Sigma| \cdot |\mathrm{dir}_{\Sigma}|)^{k-1} \cdot |\Sigma| + 2^{Q_{\mathcal A}}$.
For the winning player, a positional winning strategy can be determined in linear time in the size of $G_{\mathcal A}^k$ (see Theorem 3.1.2\ in \cite{Graedel03}, which can easily be adapted to safety games).
For a positional winning strategy of \Out, the above construction yields a DTDT with delay bounded to $k$ that uses at most $Q_{\mathcal A} \cdot (|\Sigma| \cdot \mathrm{dir}_{\Sigma})^{k-1}$ states.

We now show the other direction.

\begin{lemma}\label{lemma:BoundedDelayTDTtoStrategy}
 If $R$ has a uniformization by a DTDT in which the output delay is bounded to $k$, then \Out\ has a winning strategy in $\mathcal G_{\mathcal A}^k$.
\end{lemma}

\ifproofs
\begin{proof}
 Assume that $R$ has a uniformization by some DTDT $\mathcal T = (Q,\Sigma,\Gamma,q_0,\Delta)$ in which the output delay in bounded to $k$.
 A winning strategy for \Out\ basically takes the moves corresponding to the output sequence that $\mathcal T$ produces for a read input sequence induced by the moves of \In.
 We construct the strategy inductively.
 In a play, a vertex $\bigl(q,y\bigr)$ is reached by a sequence of moves that describe a labeled path $xiy \in \mathrm{Path}_\Sigma$ with $x, y \in \mathrm{Path}_\Sigma$, and $i \in \mathrm{dir}_{\Sigma}$.
 Let $path(xi) = u$ and $path(xiy) = v$.
 The strategy in $G_{\mathcal A}^k$ can be chosen such that in every play according to the strategy for each reached vertex $\bigl(q,y\bigr)$ the following property is satisfied.
 There is a $t \in T_{\Sigma}^{xiy}$ such that the deterministic run $\rho_{\mathcal A}$ of $\mathcal A$ on $t \otimes \mathcal T(t)$ yields $\rho_{\mathcal A}(u) = q$.
 Further, if $||y|| > 1$, then there is a configuration $(t,t',\varphi)$ of $\mathcal T$ with $(t,q_0,\varphi_0) \rightarrow^*_{\mathcal T} (t,t',\varphi)$ reachable such that there is $u \in D_{t'}$ with $\varphi(u) = v$, or there is $u' \in D_{t'}$ with $u \sqsubset u'$ and $\varphi(u') = vj$ for some $j \in \mathrm{dir}_{\Sigma}$.

 We define the strategy as follows.
 First, we consider the case $y = f \in \Sigma$, i.e., $||y|| = 1$ and $u = v$.
 For a $t \in T_{\Sigma}^{xif}$ let $s = t[\circ/u] \cdot f$ be the tree that is obtained by deleting all nodes below $u$.
 Then, \Out\ can make her next move according to the outcome of $s \otimes \mathcal T(s)$ at node $u$.
 If output was produced that is mapped to $u$, then there has to exist a rule $r$ of the form $(q,\val{t \otimes \mathcal T(t)}(u),q_1,\dots,q_n) \in \Delta_{\mathcal A}$ since $\mathcal T$ uniformizes $R$.
 Also, there has to exist an outgoing $r$-edge from $(q,f)$ that \Out\ can take.
 Otherwise, if no output was produced that is mapped to $u$, then the output is dependent on a node below $u$.
 Thus, it must hold that there is a configuration $(t,t',\varphi)$ with $\varphi(u) = v$ reachable and there exists a successor configuration $(t,t'',\varphi')$ with $\varphi'(u) = vj$ for some $j \in \mathrm{dir}_{\Sigma}$.
 \Out\ delays the output and chooses direction $j$ as next move.

 Secondly, we consider the case $||y|| > 1$.
 There are two possibilities.
 Either $\mathcal T$ reaches a configuration $(t,t',\varphi)$ with $\varphi(u) = v$ and produces no output in the next configuration step, or $\mathcal T$ reaches  $(t,t',\varphi)$ with $\varphi(u') = vj$ for some $u'$ with $u \sqsubset u'$ and some $j \in \mathrm{dir}_{\Sigma}$, meaning that $\mathcal T$ has produced output after reading the input symbol at node $v$.
 If no output was produced, \Out\ also delays.
 This can happen at most $k$ times in a row, since the output delay in $\mathcal T$ is bounded to $k$.
 Otherwise, since $\mathcal T$ uniformizes $R$, there has to exist a rule $r$ of the form $(q,\val{t \otimes \mathcal T(t)}(u),q_1,\dots,q_n) \in \Delta_{\mathcal A}$.
 We show that there also exists an $r$-labeled edge outgoing from $(q,y)$ that \Out\ can choose.
 We have to prove that the $r$-edge constraints of type iv) are satisfied.
 Let $\val{t \otimes \mathcal T(t)}(u) = (f,g) \in \Sigma \times \Gammabot$.
 Consider $l \neq j$ with $l \in \{1,\dots,n\}$.
 There are three possibilities for $l$.
 First, if $l \leq rk(f), rk(g)$, then $t_l \otimes \mathcal T(t)|_{ul} \in T(\mathcal A_{ql})$ for all $t_l \in T_{\Sigma}$.
 Assume that is not true, then there exists $t_l' \in T_{\Sigma}$ such that $t_l' \otimes \mathcal T(t)|_{ul} \notin T(\mathcal A_{ql})$.
 Since $\dom{t|{ul}} \cap \dom{t|v} = \emptyset$, we can pick $t' = t[\circ/ul] \cdot t_l'$ and obtain $\mathcal T(t) = \mathcal T(t')$.
 Thus, $t_l' \otimes \mathcal T(t)|_{ul} \in T(\mathcal A_{ql})$ which is a contradiction.
 Secondly, if $rk(g) < l \leq rk(f)$, then $t_l \otimes \bot \in T(\mathcal A_{ql})$ for all $t_l \in T_{\Sigma}$ and thirdly, if $rk(g) < l \leq rk(f)$, then $\bot \otimes \mathcal T(t)|_{ul} \in T(\mathcal A_{ql})$.
 For the correctness, the same argumentation as in the case $l \leq rk(f), rk(g)$ can be applied.

 As we have seen, \Out\ never reaches a vertex without outgoing edges and therefore wins.
\end{proof}
\fi

As a consequence of Lemma \ref{lemma:StrategytoBoundedDelayTDT} and Lemma \ref{lemma:BoundedDelayTDTtoStrategy} together with the fact that a winning strategy for \Out\ can effectively be computed in $\mathcal G_{\mathcal A}^k$ we immediately obtain Theorem \ref{theorem:BoundedDelayDecidableTDT}.

\section{Unbounded Delay}
\label{sec:4}
Previously, we considered the question whether there exists a uniformization without input validation of a D$\downarrow$TA-recognizable relation with D$\downarrow$TA-recognizable domain such that the output delay is bounded.
In this section, we will show, that this question is also decidable if the output delay is unbounded.
Similar to \cite{uniformizationAutomataTheory} for automatic word relations, we will see that if the output delay exceeds a certain bound, then we can decide whether the uniformization is possible or not.

The intuition is that if it is necessary to have such a long delay between input and output, then only one path in the tree is relevant to determine an output tree.
We can define this property by introducing the term path-recognizable function.
If a relation is uniformizable by a path recognizable function, then the relation has a uniformization by a DTDT that first deterministically reads one path of the input tree and then outputs a matching output tree.

Formally, we say a relation $R$ is \textit{uniformizable by a path-recognizable function}, if there exists a DTDT $\mathcal T$ that uniformizes $R$ such that $\Delta_{\mathcal T}$ only contains transitions of the following form:
\begin{center}
 $q(f(x_1,\dots,x_i)) \rightarrow q'(x_{j_1})$\quad or\quad $q(a) \rightarrow t$,
\end{center}
where $f \in \Sigma_i$, $i > 0$, $a \in \Sigma_0$, $q,q' \in Q$ and $j_1 \in \{1,\dots,i\}$ and $t \in T_{\Gamma}$.

In the following, we will show that there exists a bound on the output delay that we have to consider in order to decide whether a uniformization by a path-recognizable function is possible.

Beforehand, we need to fix some notations.
For $R$, $\pi \in \mathrm{Path}_{\Sigma}$ and $q \in Q_{\mathcal A}$ let 
\begingroup
\setlength{\abovedisplayskip}{.5\columnsep }
\setlength{\belowdisplayskip}{.5\columnsep }
\begin{equation*}
 R^{\pi} := \{ (t,t') \in R \mid t \in T_{\Sigma}^{\pi}\} \text{ and } R^{\pi}_q := \{ (t,t') \in R(\mathcal A_q) \mid t \in T_{\Sigma}^{\pi}\}.
\end{equation*}
\endgroup
If $q = q_0^{\mathcal A},$ then $R^{\pi}_{q}$ corresponds to $R^{\pi}$, if additionally $\pi = \varepsilon$, then $R^{\pi}_{q}$ corresponds to $R$.
Note, a D$\downarrow$TA that recognizes $R_q^\pi$ can be easily constructed from $\mathcal A$.

Since we will consider labeled paths through trees, it is convenient to define the notion of convolution also for labeled paths.
For a labeled path $x \in \mathrm{Path}_{\Sigma}$ with $||x|| > 0$, let $\dom{x} := \{ u \in \mathrm{dir}_{\Sigma}^* \mid u \sqsubseteq path(x)\}$ and $\val{x} : \dom{x} \rightarrow \Sigma$, where $\val{x}(u) = lbls(x)[i]$ if $u \in \dom{x}$ with $|u| = i+1$.
Let $x \in \mathrm{Path}_{\Sigma}$, $y \in \mathrm{Path}_{\Gamma}$ with $path(y) \sqsubseteq path(x)$ or $path(x) \sqsubseteq path(y)$, then the \textit{convolution} of $x$ and $y$ is $x \otimes y$ defined by $\dom{x \otimes y} = \dom{x} \cup \dom{y}$, and $\val{x \otimes y}(u) = (\val{x}^\bot(u),\val{y}^\bot(u))$ for all $u \in \dom{x \otimes y}$, where $\val{x}^\bot(u) = \val{x}(u)$ if $u \in \dom{x}$ and $\val{x}^\bot(u) = \bot$ otherwise, analogously defined for $\val{y}^\bot(u)$.

Furthermore, it is useful to relax the notion of runs to labeled paths.
Let $i \in \mathrm{dir}_{\Sigma}$, $x \in \mathrm{Path}_{\Sigma}$, $y \in \mathrm{Path}_{\Gamma}$ such that $x \otimes y$ is defined, i.e., $path(y) \sqsubseteq path(x)$ or $path(x) \sqsubseteq path(y)$.
We define $\rho_{\mathcal A}: \mathrm{dir}_{\Sigma}^* \rightarrow Q_{\mathcal A}$ to be the partial function with $\rho_{\mathcal A}(\varepsilon) = q_0^{\mathcal A}$, and for each $u \in \dom{x \otimes y}$: if $q := \rho_{\mathcal A}(u)$ is defined and there is a transition $(q,\val{x \otimes y}(u),q_1,\dots,q_i) \in \Delta_{\mathcal A}$, then $\rho_{\mathcal A}(u.j) = q_j$ for all $j \in \{1,\dots, i\}$.
Let $path(x \otimes y) = v$.
Shorthand, we write 
\begingroup
\setlength{\abovedisplayskip}{.5\columnsep }
\setlength{\belowdisplayskip}{.5\columnsep }
\begin{equation*}
 \mathcal A: q_0^\mathcal A \xrightarrow{x \otimes y}_i q,
\end{equation*}
\endgroup
if $q := \rho_{\mathcal A}(vi)$ is defined.
We write $\mathcal A: q_0^\mathcal A \xrightarrow{x \otimes y} F_{\mathcal A}$ if $(\rho_{\mathcal A}(v),\val{x \otimes y}(v)) \in \Delta_{\mathcal A}$ to indicate that the (partial) run $\rho_{\mathcal A}$ of $\mathcal A$ on $x \otimes y$ is accepting.

Sometimes it is sufficient to consider only the output that is mapped to a certain path.
For an input tree $t \in T_{\Sigma}$ or $t \in S_{\Sigma}$ and a path $u \in \mathrm{dir}_{\Sigma}^*$, we define
\begingroup
\setlength{\abovedisplayskip}{.5\columnsep }
\setlength{\belowdisplayskip}{.5\columnsep }
\begin{equation*}
 out_{\mathcal T}(t,u) := \{ \pi \in \mathrm{Path}_{\Gamma} \mid \mathcal T(t) \in T_{\Gamma}^{\pi} \text{ and } (path(\pi) \sqsubseteq u \text{ or } u \sqsubseteq path(\pi)) \}
\end{equation*}
\endgroup
to be the set of labeled paths through the output tree $\mathcal T(t)$ along $u$.
Note, that if $\mathcal T$ is deterministic and $||\mathcal T(t)||^u < |u|$, then $out_{\mathcal T}(t,u)$ is a singleton set.

We introduce a partial function that yields the state transformations on a path $\pi$ induced by the input sequence of $\pi$ together with some output sequence on the same path of same or smaller length.
Formally, for $x \in \mathrm{Path}_{\Sigma}$, $y \in \mathrm{Path}_{\Gamma}$ and a direction $i \in \mathrm{dir}_{\Sigma}$ such that $path(y) \sqsubseteq path(x)$, we define the partial function $\tau_{xi,y} : Q_{\mathcal A} \rightarrow Q_{\mathcal A}$ with $\tau_{xi,y}(q) := q'$ if $\mathcal A: q \xrightarrow{x \otimes y}_i q'$ and for each $uj$ with $u \in \dom{x}: uj \not \sqsubseteq path(xi)$ and $j \in \{1,\dots,rk\bigl((\val{x}^\bot(u),\val{y}^\bot(u))\bigr)\}$ holds
 \begin{itemize}
  \item if $r := \rho_{\mathcal A}(uj)$ and $j \leq rk(\val{x}^\bot(u))$, then there exists $t' \in T_{\Gamma}$ such that for all $t \in T_{\Sigma}$ holds $t \otimes t' \in T(\mathcal A_r)$, and
  \item if $r := \rho_{\mathcal A}(uj)$ and $j > rk(\val{x}^\bot(u))$, then there exists $t' \in T_{\Gamma}$ such that $\bot \otimes t' \in T(\mathcal A_r)$,
 \end{itemize}
where $\rho_{\mathcal A}$ is the run of $\mathcal A_q$ on $x \otimes y$.
Lemma \ref{lemma:edgecontraints} implies that it is decidable whether $\tau_{xi,y}(q)$ is defined.
Basically, if $q' := \tau_{xi,y}(q)$ is defined, then there exists a fixed (partial) output tree $s' \in S_{\Gamma}^{yi\circ}$ such that for each input tree $t \in T_{\Sigma}^x$ there exists some $t' \in T_{\Gamma}$ such that $t \otimes (s' \cdot t') \in T(\mathcal A_q)$. 

We define the profile of a labeled path segment $xi$ to be the set that contains all possible state transformations induced by $x$ together with some $y$ of same or smaller length.
Formally, let $x \in \mathrm{Path}_{\Sigma}$ and $i \in \mathrm{dir}_{\Sigma}$, we define the profile of $xi$ to be $P_{xi} = (P_{xi,=},P_{xi,<},P_{xi,\varepsilon})$ with
 \begin{itemize}
  \item $P_{xi,=} := \{\tau_{xi,y} \mid |y| = |x| \}$
  \item $P_{xi,<} := \{\tau_{xi,y} \mid y \neq \varepsilon \text{ and } |y| < |x|\}$
  \item $P_{xi,\varepsilon} := \{\tau_{xi,y} \mid y = \varepsilon\}$.
 \end{itemize}
A segment $xi \in (\Sigma\mathrm{dir}_{\Sigma})^*\mathrm{dir}_{\Sigma}$ of a labeled path is called idempotent if $P_{xi} = P_{xixi}$.

As a consequence of Ramsey's Theorem \cite{ramsey1930problem}, we obtain the next remark.

\begin{remark}\label{remark:ramsey}
 There exists a bound $K \in \mathbbm{N}$ such that each labeled path $\pi \in \mathrm{Path}_{\Sigma}$ with $||\pi|| \geq K$ contains an idempotent factor.
\end{remark}

For the rest of this paper we fix how we repeat the part of a tree that contains an idempotent factor in a labeled path segment.
Let $x,y \in \mathrm{Path}_{\Sigma}$, $i,j \in \mathbbm{N}$ with $y \neq \varepsilon$ and $yj$ idempotent.
For any $t \in T_{\Sigma}^{xiy}$ we fix $t^n$ to be the tree that results from repeating the idempotent factor $n$ times.
More formally, let $path(x) = u$ and $path(y) = v$, we define
\begingroup
\setlength{\abovedisplayskip}{.5\columnsep }
\setlength{\belowdisplayskip}{.5\columnsep }
\begin{equation*}
t^n := \underbrace{t[\circ/ui]}_{s_x} \cdot (\underbrace{t|_{ui}[\circ/uivj]}_{s_y})^n \cdot \underbrace{t|_{uivj}}_{\hat{t}} \text{ for $n \in \mathbbm{N}$.}
\end{equation*}
\endgroup

The following Lemma shows that is it decidable whether a relation has a uniformization by a path-recognizable function. 

\begin{lemma}\label{lemma:decidablePathrecfunction}
 For $q \in Q_{\mathcal A}$, $x,y \in \mathrm{Path}_{\Sigma}$, $i,j \in \mathbbm{N}$ with $path(x) = u$, $path(y) = v$, $y \neq \varepsilon$ and $yj$ idempotent, it is decidable whether $R^{xiy}_q$ can be uniformized by a path-recognizable function.
\end{lemma}

The following Lemma establishes the connection between long output delay and path-recognizable functions.
Basically, the lemma states that if there exists a uniformization by a DTDT such that an idempotent path segment can be repeated any number of times and the length of the output on the repetition is bounded, i.e., the output delay is unbounded, then there also exists a uniformization by a path-recognizable function.

\begin{lemma}\label{lemma:unifTDTtopathrecfunction}
 Let $q \in Q_{\mathcal A}$, $x,y \in \mathrm{Path}_{\Sigma}$, $i,j \in \mathbbm{N}$ with $path(x) = u$, $path(y) = v$, $y \neq \varepsilon$ and $yj$ idempotent.
 If $R^{xiy}_q$ is uniformized by a DTDT $\mathcal T$ such that $||\mathcal T(s_x \cdot s_y^n)||^{ui(vj)^n} \leq |ui|$ for each $t \in T_{\Sigma}^{xiy}$ and for each $n \in \mathbbm N$, then $R^{xiy}_q$ can be uniformized by a path-recognizable function.
\end{lemma}

As we have seen, if a transducer that uniformizes a relation introduces long output delay, then the relation can also be uniformized by a path-recognizable function.

Now that we have completed all preparations, we present a decision procedure for the case of unbounded delay.
Therefore, we consider a similar safety game as in the previous section on uniformization with bounded output delay.
We only have to adapt the game graph if the input sequence is ahead $K$ steps.
Let $\mathcal G_{\mathcal A}^K$ denote the modified game.
From each vertex $\left(q,\pi\right) \in \Vout$ with $||\pi|| = K$ we add a move that allows \Out\ to stay in this vertex if there exists a factorization of $\pi = xiyjz$ with $x,y,z \in \mathrm{Path}_{\Sigma}$, $i,j \in \mathrm{dir}$ and $yj$ is idempotent such that $R^{xiy}_q$ can be uniformized by a path-recognizable function without input validation.
These changes to the game graph can be made, because if the input is $K$ steps ahead, then there exists a factorization of the input sequence that contains an idempotent factor and Lemma \ref{lemma:decidablePathrecfunction} implies that it is decidable whether there exists a corresponding uniformization by a path-recognizable function.
 
\begin{lemma}\label{lemma:unboundedDelay}
 $R$ has a uniformization if, and only if, \Out\ has a winning strategy in the safety game $\mathcal G_{\mathcal A}^K$.
\end{lemma}

\ifproofs
\begin{proof}
 Assume that \Out\ has a winning strategy in $\mathcal G_{\mathcal A}^K$, then there also exists a positional winning strategy for \Out.
 To construct a DTDT $\mathcal T$ that uniformizes $R$, we proceed as presented in the proof of Lemma \ref{lemma:StrategytoBoundedDelayTDT} with one addition.
 We construct for each $\left(q,\pi\right) \in \Vout$ such that $||\pi|| = K$ and there is $\pi = xiyjz$ such that $R^{xiy}_q$ can be uniformized by a path-recognizable function, a DTDT $\mathcal T_q^{xiy}$ that uniformizes $R^{xiy}_q$.
 In $\mathcal T$ we switch to $\mathcal T_q^{xiy}$ at the respective states.
 
 For the other direction, assume that $R$ is uniformized by some DTDT $\mathcal T$.
 Again, the proof is similar to the proof of Lemma \ref{lemma:BoundedDelayTDTtoStrategy}.
 Thus, we only describe how the strategy is chosen if the output delay in $\mathcal T$ exceeds $K$.
 If the play reaches a vertex  $\left(q,\pi\right) \in \Vout$ with $||\pi|| = K$, there is a factorization of $\pi = xiyjz$ with $x,y,z \in \mathrm{Path}_{\Sigma}$, $i,j \in \mathrm{dir}$ such that $yj$ is idempotent.
 Let $path(x) = u, path(y) = v$ and $path(z) = w$.
 Let $\mathcal T_s$ uniformize $R_q$ and pick any $t \in T_{\Sigma}^{\pi}$, if $||out_{\mathcal T_s}(t^n = s_x \cdot s_y^n \cdot \hat{t},ui(vj)^n)|| < ui(vj)^n$ for all $n \in \mathbbm{N}$, then Lemma \ref{lemma:unifTDTtopathrecfunction} implies that $R^{xiy}$ can be uniformized by a path-recognizable function.
 In this case, \Out\ stays in this vertex from then on and wins.
 
 Otherwise, there exists $m \in \mathbbm{N}$ such that $||out_{\mathcal T_s}(s_x \cdot s_y^m \cdot \hat{t},ui(vj)^m)|| \geq ui(vj)^m$.
 Consider the factorization of $out_{\mathcal T_s}(s_x \cdot s_y^m \cdot \hat{t},ui(vj)^m) = o_1io_2jo_3$ such that $|o_1i| = |xi|$ and $|o_2j| = (yj)^m$.
 Since $yj$ is idempotent we can choose some $o$ of length $K$ such that $\mathcal A: q \xrightarrow{xiy \otimes o_1o}_j q'$ and $\mathcal A: q \xrightarrow{xi(yj)^{m-1}y \otimes o_1o_2}_j q''$ with $q' = q''$.
 Then \Out\ makes $K$ moves according to $o$ leading to some $\left(q',z\right) \in \Vout$.
 From there, \Out\ takes the transitions according to $o_3$.
\end{proof}
\fi

As a consequence of Lemma \ref{lemma:unboundedDelay} and the fact that a winning strategy for \Out\ in $\mathcal G_{\mathcal A}^K$ can effectively be computed we immediately obtain our main result.

\begin{theorem}
 It is decidable whether a D$\downarrow$TA-recognizable relation with total domain has a uniformization by a deterministic top-down tree transducer.
\end{theorem}

As mentioned in the beginning, the presented results are also valid for D$\downarrow$TA-recognizable relations with D$\downarrow$TA-recognizable domain in the sense that a TDT that realizes a uniformization of a relation may behave arbitrarily on trees that are not part of the domain.
The presented constructions have to be adapted such that \In, given a D$\downarrow$TA for the domain, also keeps track of the state in the input tree in order to play only correct input symbols.

\section{Input Validation}
\label{sec:5}
In the former section, we assumed that a top-down tree transducer that implements a uniformization of a relation is only given valid input trees.
In this section we consider the case that a top-down transducer also has to validate  the correctness of a given input tree.

We will see that in this case it can be necessary that a transducer takes divergent paths for input and output.
The following example shows that there exists a D$\downarrow$TA-recognizable relation with D$\downarrow$TA-recognizable domain that can be uniformized by a DTDT, but every such DTDT has a reachable configuration $(t,t',\varphi)$ such that $\varphi(u) \not\sqsubseteq u$ and $u \not\sqsubseteq \varphi(u)$ for some node $u$.

\begin{example}\label{ex:inputoutputDivergent}
 Let $\Sigma$ be given by $\Sigma_2 = \{f\}$ and $\Sigma_0 = \{a,b\}$.
 We consider the relation $R_1 \subseteq T_{\Sigma} \times T_{\Sigma}$ defined by $\{\bigl(f(b,t),f(t',b)\bigr) \mid \neg \exists u \in \dom{t}: \val{t}(u) = b\}$.
 Clearly, both $R_1$ and $dom(R_1)$ are D$\downarrow$TA-recognizable.
 Intuitively, a DTDT $\mathcal T$ that uniformizes $R_1$ must read the whole right subtree $t|_2$ of an input tree $t$ to verify that there is no occurrence of $b$.
 If an $f$ in $t|_2$ is read and no output is produced, a DTDT can either continue to read left or right, but cannot verify both subtrees.
 Therefore, in order to verify $t|_2$, a DTDT has to produce an output symbol at each read inner node which results in an output tree of the same size.
 Clearly, the relation $R_1$ is uniformized by the following DTDT $\mathcal T = (\{q_0,q_1,q_2\},\Sigma,\Sigma,q_0,\Delta)$ with $\Delta =$
 \begingroup
 \setlength{\abovedisplayskip}{.5\columnsep }
 \setlength{\belowdisplayskip}{.5\columnsep }
 \begin{align*}
  \{ \quad q_0(f(x_1,x_2)) & \rightarrow f(q_1(x_2),q_2(x_1)),\quad q_1(a) \rightarrow b,\\
  q_1(f(x_1,x_2)) & \rightarrow f(q_1(x_1),q_1(x_2)),\quad q_2(b) \rightarrow b \quad \}.
 \end{align*}
 \endgroup
 However, there exists no DTDT $\mathcal T'$ that uniformizes $R_1$ such that the read input sequence and the produced output are on the same path.
 Assume such a DTDT $\mathcal T'$ exists, then for an initial state $q_0$ there is a transition of the form $q_0(f(x_1,x_2)) \rightarrow f(q_1(x_1),q_2(x_2))$.
 It follows that $\mathcal T'_{q_2}$ must induce the relation $ \{ (t,b) \mid t \in T_ {\Sigma} \wedge \neg \exists u \in \dom{t}: \val{t}(u) = b\}$.
 The only output that $\mathcal T'_{q_2}$ can produce is exactly one $b$.
 Thus, there is a transition with left-hand side $q_2(f(x_1,x_2))$ that has one of the following right-hand sides: $b$, $q_3(x_1)$, or $q_3(x_2)$.
 No matter which right-hand side is chosen, $dom(R(\mathcal T'_{q_2}))$ must also contain trees with occurrences of $b$.
\end{example}

It follows directly from the above example that the presented decision procedure is invalid if the domain of a considered relation is not total.
However, if we restrict ourselves to uniformizations such that a DTDT only contains rules of the form $q(f(x_1,\dots,x_i)) \rightarrow w[q_1(x_{j_1}),\dots,q_n(x_{j_n})]$, where $w \in \Gamma(X_n)$ i.e., read input symbol and correspondingly produced output begin always on the same tree level, it is possible to adapt the presented decision procedure from Section \ref{sec:3}.
We refer to this kind of DTDTs as DTDTs without delay.

\begin{theorem}
 It is decidable whether a D$\downarrow$TA-recognizable relation with D$\downarrow$TA-recognizable domain has a uniformization by a deterministic top-down tree transducer without delay.
\end{theorem}

For this purpose, we can change the game graph in the following way.
Let $\mathcal A$ be a D$\downarrow$TA for a relation and $\mathcal B$ be a D$\downarrow$TA for its domain.
The main differences to the previous section is that the vertices in the game graph keep track of the current state of $\mathcal B$ on the input sequence played by \In\ and keep track of the state of $\mathcal A$ on the combined part of all possible input sequences and the current output sequence of \Out\ which is not necessarily the same as the input sequence played by \In.
The move constraints for \Out\ will be chosen such that it is guaranteed that the input sequence is valid, and the combined part of all possible input sequences together with her output sequence is valid.
Details for this construction can be found in \cite{masterarbeit}.

\section{Conclusion}
\label{sec:6}
In this paper, we focused on synthesis of deterministic top-down tree transducers from deterministic top-down tree automaton-definable specifications.
We have shown that is decidable whether a specification can be realized by a top-down tree transducer under the restriction that the transducer is not required to validate the input, meaning that a transducer implementing a uniformization can behave arbitrarily on invalid inputs.
If uniformization is possible, our decision procedure yields a top-down tree transducer that realizes the specification.

We have seen that the presented decision procedure concerning uniformization without input validation cannot be transferred directly to decide the problem corresponding to the classical uniformization question (with input validation).
The reason for this is that in the employed transducer model it is not possible to verify the input without producing output.

In the future, we want to investigate synthesis of tree transducers from general non-deterministic tree relations.
It looks promising to use guidable tree automata \cite{loedinghabil} for the specifications.
It seems that the presented decision procedure remains valid in case of uniformization without input validation.

% ---------- References ----------

%\nocite{*}
\bibliographystyle{eptcs}
\bibliography{paper}

% ---------- Appendix ----------

\theoremstyle{empty}
\theorembodyfont{\slshape}
\theoremsymbol{}
\renewtheorem{theorem}{Theorem}
\renewtheorem{lemma}{Lemma}
\renewtheorem{remark}{Remark}

\ifappendix
\newpage
\section*{Appendix}
\label{sec:appendix}
\input{appendix.tex}
\fi
\end{document}